\documentclass[12pt]{article}
\pdfoutput=1
\usepackage[a4paper,text={16.8cm,22.4cm}]{geometry}
\PassOptionsToPackage{dvipsnames}{xcolor}
\usepackage{hf-tikz,amsmath,amsfonts,braket,slashed,amssymb,bm,psfrag,graphicx,color,colortbl,dsfont,euscript,ctable,bbm,hyperref} 
\usepackage[small,labelfont=bf]{caption}
\usetikzlibrary{arrows.meta}
\RequirePackage[sort&compress,square,comma,numbers]{natbib}
\usepackage{stmaryrd} 
\usepackage{comment}
\usepackage{soul}
\allowdisplaybreaks
\renewcommand{\arraystretch}{1.3}
\newcommand{\spac}{{\hspace{0.3mm}}}
\newcommand{\nsl}{\rlap{\hspace{0.25mm}/}{n}}
\newcommand{\nbsl}{\rlap{\hspace{0.25mm}/}{\nb}}
\newcommand{\nb}{{\bar n}}
\newcommand{\vsl}{\rlap{\hspace{0.25mm}/}{v}}

\newcommand{\om}{{\omega}}
\newcommand{\order}[1]{\mathcal{O}\!\left(#1\right)}
\newcommand{\myR}{R^{(M_2,B)}}

\definecolor{LightGray}{rgb}{0.85,0.85,0.85}

\definecolor{AccentColor}{rgb}{0.0,0.47,.90}
\newcommand{\shadecolor}[1]{AccentColor!#1!white}
\numberwithin{equation}{section}
\begin{document}

\begin{titlepage}

\begin{flushright}
\normalsize
MITP-26-040\\
September 1, 2026
\end{flushright}

\vspace{0.4cm}
\begin{center}
\LARGE\bf
Electromagnetic tails of {\boldmath $B$}-meson light-cone distribution amplitudes
\end{center}

\vspace{0.4cm}
\begin{center}
\textsc{Max Ferr\'e\spac$^a$ and Matthias Neubert\spac$^{a,b}$}

\vspace{6mm}
\textsl{${}^a$PRISMA$^{++}$ Cluster of Excellence \& Mainz Institute for Theoretical Physics\\
Johannes Gutenberg University, Staudingerweg 9, D-55128 Mainz, Germany\\[0.3cm]
${}^b$Department of Physics \& LEPP, Cornell University, Ithaca, NY 14853, U.S.A.}
\end{center}
\vspace{0.8cm}

\begin{abstract}
We investigate electromagnetic corrections to the $B$-meson decay constant and light-cone distribution amplitudes (LCDAs) within the framework of heavy-quark effective theory. The inclusion of QED effects extends the support of the LCDAs to negative values of the light-cone momentum. We clarify the relation between this region and QED-generalized decay constants sensitive to electromagnetic final-state radiation. We derive model-independent expressions for the asymptotic behavior in the large negative-momentum region of the QED-generalized LCDAs $\Phi_\pm(\omega,\mu)$ in terms of integrals over the two- and three-particle LCDAs defined in pure QCD. We show that our result for $\Phi_+$ is consistent with the renormalization-group equation satisfied by this object and study its evolution using the known one-loop anomalous dimension. Finally, we propose a phenomenological model consistent with these constraints.
\end{abstract}

\end{titlepage}

\tableofcontents

\section{Introduction}

Exclusive decays of heavy mesons provide a powerful probe of the flavor structure of the Standard Model (SM) and offer a sensitive window into possible effects of new physics, particularly in rare decays that are suppressed within the SM. Processes involving energetic light particles in the final state, however, require a precise understanding of the interplay between perturbative and non-perturbative QCD dynamics. Within this context, light-cone distribution amplitudes (LCDAs) play a central role, as they encode the momentum distribution of the soft spectator quark inside the meson and enter factorization theorems for a large class of exclusive decay amplitudes.

For the $B$ meson, the leading- and subleading-power LCDAs have been extensively studied in the framework of heavy-quark effective theory (HQET) \cite{Grozin:1996pq,Beneke:2000wa,Kawamura:2001jm,Lange:2003ff,Lee:2005gza,Braun:2017liq}. These non-perturbative objects appear ubiquitously in QCD factorization and soft-collinear effective theory descriptions of heavy-to-light transitions, including radiative, semileptonic, and non-leptonic $B$-meson decays. Their properties under renormalization-group (RG) evolution, their asymptotic behavior at large positive light-cone momentum, and the role of higher-power contributions have all been investigated in considerable detail. In particular, perturbative QCD corrections generate a characteristic radiative tail at large positive momentum, whose structure can be derived in a model-independent way using a local operator product expansion (OPE) \cite{Lee:2005gza}. More recently, analogous radiative tails have been studied for multi-particle heavy-hadron distribution amplitudes, including three-particle $\Lambda_b$ LCDAs \cite{Feldmann:2025dcs} and $B$-meson di-light-cone distribution amplitudes involving two distinct light-like directions \cite{Bartocci:2026lhe}. These studies illustrate how short-distance singularities of non-local HQET operators generate power-law radiative tails in momentum space.

With the increasing precision of experimental measurements, electromagnetic effects can no longer be neglected in many phenomenologically relevant observables, ushering in an era of precision theoretical predictions that require systematic studies of QED corrections. Soft-photon radiation induces infrared-sensitive effects that are intrinsically tied to the charges and directions of the external particles involved in the process. Unlike soft gluons, soft photons are not confined and can couple to external particles, implying that hadronic quantities defined in the presence of QED are no longer universal. This feature has recently attracted significant attention in the context of exclusive weak decays and precision flavor observables \cite{Beneke:2019slt,Beneke:2020vnb,Beneke:2022msp,Cornella:2022ubo,Cornella:2026lkp}. In particular, it was shown that electromagnetic corrections modify the analytic structure of the $B$-meson LCDAs and generate support for negative values of the light-cone momentum variable \cite{Beneke:2022msp}. The appearance of such negative-momentum contributions reflects the non-local nature of soft-photon interactions, which couple to charged external particles through eikonal Wilson lines. As a consequence, the standard definitions of $B$-meson decay constants and transition form factors in terms of hadronic matrix elements of local quark currents need to be extended in subtle ways in the presence of QED effects. A consistent treatment requires introducing generalized hadronic quantities, which explicitly account for electromagnetic final-state radiation and the associated subtraction of soft-collinear overlap contributions. 

In this work, we investigate electromagnetic corrections to the $B$-meson decay constant and LCDAs in the framework of HQET. We focus on the structure of the QED-induced negative-momentum tail and its relation to generalized decay constants sensitive to soft final-state radiation. As our most important result, we derive model-independent expressions for the asymptotic behavior of the QED-generalized LCDAs $\Phi_\pm(\omega,\mu)$ at large negative momentum $\omega\ll-\Lambda_{\rm QCD}$ in terms of integrals over the two- and three-particle LCDAs defined in pure QCD. We also study the RG evolution of the negative branch of the leading-order LCDA and show that our result is consistent with the anomalous dimension derived in \cite{Beneke:2022msp}. Building on these findings, we construct a phenomenological model for the negative-support component of the leading-power LCDA consistent with the asymptotic constraints and RG properties.

\section{QED generalizations of decay constants and LCDAs}
\label{sec:LCDA}

Working in the framework of HQET, we denote by $v^\mu$ the 4-velocity of the $B$-meson (with $v^2=1$) and  describe the $b$ quark via the heavy-quark field $b_v(x)$ satisfying  
\begin{equation}
   \vsl\spac b_v = b_v \,, \qquad i
   v\cdot D\,b_v = 0 \,.
\end{equation}
We define a generic bilocal current  $\bar{u}(zn)[zn,0]\spac\Gamma\spac b_v(0)$ consisting of a soft spectator quark field $u$ and the HQET field $b_v$, separated by a light-like distance $z\spac n^\mu$. Gauge invariance is ensured by the insertion of a finite-length Wilson line $[zn,0]$ connecting the two spacetime points. To parameterize the light-cone separation, we introduce two light-like reference vectors $n^\mu$ and $\nb^\mu$ satisfying
\begin{equation}
   n^2 = \nb^2 = 0 \,, \qquad 
   n\cdot\nb = 2 \,, \qquad 
   v^\mu = v\cdot\nb\,\frac{n^\mu}{2} + v\cdot n\,\frac{\nb^\mu}{2} \,.
\end{equation}
For a generic Dirac matrix $\Gamma$, the $B$-meson matrix element of such a bilocal current can be parameterized through the leading and subleading LCDAs $\phi_+$ and $\phi_-$. Employing the trace formalism of HQET \cite{Neubert:1993mb}, and performing a Fourier transformation to momentum space, the matrix element takes the form \cite{Grozin:1996pq}
\begin{equation}\label{eq:nLCDA}
\begin{aligned}
   &\langle\spac 0\spac|\spac \bar u\,
    \delta\bigg(\om-\frac{in\cdot\!\overleftarrow D}{n\cdot v}\bigg)
    \spac\Gamma\spac b_v\spac|B^-(v)\rangle \\
   &= \frac{i}{2} \sqrt{m_B}\,F_{\rm QCD}\,\mathrm{Tr}\left\{ 
    \left[ \frac{\nbsl}{2\spac \nb\cdot v}\,\phi_{+}( \om)
    + \frac{\nsl}{2\spac n\cdot v}\,\phi_{-}(\om) \right] 
    \Gamma\,\frac{1+\slashed{v}}{2}\spac\gamma_5\spac \right\} , 
\end{aligned}
\end{equation}
where $\omega$ denotes the momentum variable conjugate to the displacement parameter $z\spac n\cdot v$. Within dimensional regularization, the local limit $z\to 0$ of this matrix elements is well defined and reduces to the matrix element of the local HQET current. The latter defines the HQET decay constant $F_{\rm QCD}$ via the relation
\begin{equation}\label{eq:FQCDdef} 
\begin{aligned}
   \langle0|\spac\bar u\spac\Gamma\spac b_v\spac|B^-(v)\rangle 
   = - \frac{i}{2} \sqrt{m_B}\spac F_{\rm QCD}\,\mathrm{Tr}\left\{ 
    \Gamma\,\frac{1+\slashed{v}}{2}\spac\gamma_5\spac \right\} . 
\end{aligned}
\end{equation}
After renormalization, the HQET decay constant is related to the $B$-meson decay constant $f_B$ via the matching relation 
\begin{equation}\label{eq:Fhqet_fB} 
   \sqrt{m_B}\spac f_B
   = \left[1 + \frac{C_F\spac\alpha_s}{4\pi} \left( 3 \ln\frac{m_b}{\mu} - 2 \right) \right] 
    F_\mathrm{QCD}(\mu) + \order{\frac{1}{m_b}} .
\end{equation}
Higher-order terms in the heavy-quark expansion have been studied in \cite{Neubert:1992fk}. Taking the local limit of the bilocal operator implies the normalization conditions
\begin{equation}\label{eq:norm}
   \int_{0}^\infty\!d\om\,\phi_{\pm}(\om) = 1 \,,
\end{equation}
which apply to the bare (dimensionally regularized) LCDAs. After renormalization, the integrals need to be defined with a cutoff $\Lambda_{\rm UV}\gg\Lambda_{\rm QCD}$, because radiative gluon exchange generates a power-law tail at large $\om$, which renders the corresponding integrals UV divergent \cite{Grozin:1996pq,Lange:2003ff}. Focusing on the leading-power two-particle LCDA $\phi_+(\om,\mu)$, and using a local OPE, one obtains \cite{Lee:2005gza}
\begin{equation}\label{eq:normUV}
   \int_{0}^{\Lambda_{\rm UV}}\!d\om\,\phi_+(\om,\mu) 
   = 1 + \frac{C_F\spac\alpha_s}{4\pi} \left[ \left( - 2 \ln^2\frac{\Lambda_{\rm UV}}{\mu} 
    + 2 \ln\frac{\Lambda_{\rm UV}}{\mu} - \frac{\pi^2}{12} \right) 
    + \frac{16\bar\Lambda}{3\Lambda_{\rm UV}} \left( \ln\frac{\Lambda_{\rm UV}}{\mu} - 1 \right) 
    \right] ,
\end{equation}
where we only show the leading perturbative and power corrections.\footnote{A consistent treatment of the LCDA in the region between $\Lambda_{\rm UV}$ and the physical upper cutoff $m_b$ requires a refactorization of $\phi_+(\omega)$ for $\omega\gtrsim\Lambda_{\rm UV}$ and a proper matching onto the LCDA defined in full QCD \cite{Beneke:2023nmj}.} The behavior of $\phi_+(\om,\mu)$ in the region $\om\gg\Lambda_{\mathrm{QCD}}$ can be derived by taking a derivative with respect to the cutoff. At one-loop order, this procedure yields the asymptotic form \cite{Lee:2005gza}
\begin{equation}\label{eq:QCDtail}
   \phi_+(\omega,\mu) \underset{\om\gg\Lambda_{\rm QCD}}{=} 
   \frac{C_F\spac\alpha_s}{\pi\spac\om} \left[ \left( \frac12 - \ln\frac{\om}{\mu} \right) 
    + \frac{4\bar\Lambda}{3\spac\om} \left( 2 - \ln\frac{\om}{\mu} \right)
    + \order{\frac{\Lambda_{\rm QCD}^2}{\om^2}} \right] .
\end{equation}
The leading $1/\om$ behavior of this expansion is entirely governed by short-distance coefficients computed perturbatively. The leading non-perturbative effects enter at subleading power through the hadronic parameter $\bar\Lambda=m_B-m_b$.

While $\phi_+$ contributes at leading power in the light-cone expansion, the  two-particle LCDA $\phi_-$ enters at next-to-leading power, i.e.\ $\order{\Lambda_{\rm QCD}/m_b}$, together with the three-particle distribution amplitudes involving an additional gluon. In particular, $\phi_-$ mixes under RG evolution with the leading three-particle LCDA denoted by $\phi_{3g}$. The latter is defined via the $B$-meson matrix element of the non-local current \cite{Kawamura:2001jm,Braun:2017liq}
\begin{align}
    \bar u(z_1n)[z_1n,z_2n]\,g_s\spac G^{\mu\nu}(z_2n)[z_2n,0]\spac\Gamma\spac b_v(0) \,,
\end{align}
which contains a soft quark field and a soft gluon field displaced along the light-cone in the $n^\mu$ direction. The fields are connected by finite-length Wilson lines, ensuring gauge invariance of the non-local operator. Matrix elements of this operator can be parameterized using the HQET trace formalism \cite{Kawamura:2001jm,Braun:2017liq}. The leading-power three-particle LCDA $\phi_{3g}$ is obtained by projecting the resulting decomposition onto an appropriate Lorentz and Dirac structure. Upon Fourier transformation to momentum space, one obtains 
\begin{equation}\label{eq:nLCDA3p}
\begin{aligned}
   &\langle0|\spac\bar u\spac\delta\bigg(\om-\frac{in\cdot\!\overleftarrow D}{n\cdot v}\bigg)
    \bigg[ \spac g_s\spac G^{\mu\nu}\spac
    \delta\bigg(\om_g-\frac{in\cdot\!\overleftarrow D}{n\cdot v}\bigg)
    \bigg]\spac n_\nu\spac\gamma_\mu^\perp\spac\frac{\nsl}{n\cdot v}\spac\gamma_5\,b_v\spac
    |B^-(v)\rangle \\
   &=- 2\sqrt{m_B}\spac F_{\rm QCD}\,\om_g\spac\phi_{3g}(\om,\om_g) \,,
\end{aligned}
\end{equation}
where the superscript ``$\perp$'' denotes components transverse to $n^\mu$ and $\nb^\mu$ and $\om,\om_g$ denote the momenta conjugate to the variables $z_1\spac n\cdot v$ and $z_2\spac\spac n\cdot v$, respectively. In the notations of  \cite{Kawamura:2001jm,Braun:2017liq} one has $\phi_{3g}(\om,\om_g)=[\Psi_A(\om,\om_g)-\Psi_V(\om,\om_g)]/\om_g=\phi_3(\om,\om_g)/\om_g$. 

In the limit of very large $\mu$, the asymptotic behavior of the renormalized LCDAs for small values of the momentum variables is dictated by conformal symmetry. One finds \cite{Braun:2017liq}
\begin{align}\label{eq:scaling}
   \phi_+(\omega,\mu) \sim \omega \,, \qquad
   \phi_-(\omega,\mu) \sim 1 \,, \qquad
   \phi_{3g}(\omega,\omega_g,\mu) \sim \omega\,\omega_g \,. 
\end{align}

In the presence of QED effects, the definitions of the decay constant and the LCDAs cease to be universal and acquire an explicit process dependence. This originates from the fact that soft photons are not confined and can thus mediate long-range interactions between electrically charged particles (hadrons and leptons) in a scattering process. A consistent generalization of the $B$-meson decay constant $F_{\rm QCD}$ probed in the leptonic decay $B^-\to\bar\nu_\mu\spac\mu^-$ leads to two form factor-like objects $F_\pm$, which are sensitive to the direction of the final-state muon \cite{Cornella:2022ubo,Cornella:2026lkp}. Likewise, a consistent generalization of $B^-\to M_1^0$ form factors leads to quantities which know about the direction of other charged final-state particles in processes such as $B^-\to M_1^0\spac M_2^-$\cite{Beneke:2020vnb}. A proper theoretical treatment of soft photon interactions requires incorporating the charges and directions of the electrically charged external states, which is naturally achieved through the introduction of Wilson lines. For the case of a single charged particle that is light and highly energetic, the couplings to soft photons can be encoded in a light-like Wilson line\footnote{For the decay $B^-\to\tau^-\spac\bar\nu_\tau$, in which the charged final-state lepton is heavy, one uses instead a time-like Wilson line along the direction set by the 4-velocity $v_\tau^\mu$  \cite{Cornella:2026ask}.} along the direction of flight $\nb^\mu$, defined as (with $Q_\mu=Q_{M_2}=Q_{B^-}=-1$)
\begin{equation}\label{eq:WLdef}
   Y_\nb^{(M_2)\dagger}(x)
   = \exp\bigg[  i\spac Q_{M_2}\spac e\!\int_0^\infty\!ds\,\nb\cdot A(x+s\nb) \bigg] \,.
\end{equation}
Including this Wilson line ensures QED gauge invariance of $b\to u$ currents.

\subsection{QED-generalized decay constants}

Starting from the matrix element of the local current in \eqref{eq:FQCDdef}, the inclusion of QED effects via the Wilson line \eqref{eq:WLdef} introduces a second reference vector $\nb^\mu$ in addition to the 4-velocity of the $B$ meson. As a result, the trace formalism allows for two reduced matrix elements $F_\pm$, the QED-generalized HQET decay constants, defined as\footnote{This definition is analogous to that introduced in \cite{Cornella:2026lkp} with the roles of $n^\mu$ and $\nb^\mu$ interchanged.}
\begin{align}\label{eq:Fdefnaive}
\begin{aligned}
   \frac{\langle\spac 0\spac|\spac\bar u\spac\Gamma\spac b_v\spac Y_\nb^{(M_2)\dagger}\spac|B^-(v)\rangle}{\myR}
   &= - \frac{i}{2}\spac\sqrt{m_B}\,\,
    \mathrm{Tr}\left\{ \left[ F_+ + \frac{\nbsl}{2\spac\nb\cdot v}\,(F_+-F_-) \right] 
    \Gamma\spac\frac{1+\slashed{v}}{2}\spac\gamma_5\spac \right\} \\
   &= \frac{i}{2}\spac\sqrt{m_B}\,\,
    \mathrm{Tr}\left\{ \left[ \frac{\nbsl}{2\spac\nb\cdot v}\,F_-
    + \frac{\nsl}{2\spac n\cdot v}\,F_+ \right] 
    \Gamma\spac\frac{1+\slashed{v}}{2}\spac\gamma_5\spac \right\} .
\end{aligned}
\end{align}
Comparison with the definition \eqref{eq:FQCDdef} valid in pure QCD shows that  
\begin{equation}
    F_\pm = F_\mathrm{QCD} + \mathcal{O}(\alpha) \,, 
\end{equation}
such that both parameters reduce to $F_\mathrm{QCD}$ in the absence of electromagnetic effects. Including QED effects, a proper definition requires removing the overlap between the soft and collinear divergences from the matrix element, which as discussed in \cite{Beneke:2019slt,Beneke:2020vnb} is accomplished by dividing the hadronic matrix element by the vacuum matrix element of two soft Wilson lines corresponding to the $B^-$ meson and the charged final-state particle, namely
\begin{equation}\label{eq:Rdef}
   \myR = \langle\spac 0\spac|\spac Y_\nb^{(M_2)\dagger}\,\overline{Y}_v^{(B^-)}\spac|\spac 0\spac\rangle \,,
\end{equation}
where
\begin{equation}
   \overline{Y}_v^{(B)}(x)
   = \exp\bigg[ i\spac Q_{B^-}\spac e\!\int_{-\infty}^0\!ds\,v\cdot A(x+sv) \bigg] \,.
\end{equation}
This procedure ensures that the resulting hadronic quantities are independent of unphysical infrared regulators. 

It was shown in \cite{Cornella:2026lkp} that the QED-generalized local current operator in \eqref{eq:Fdefnaive} mixes with non-local current operators under renormalization, a feature that is highly unusual. Moreover, it was explained in this reference that a consistent scale factorization for the leptonic decay $B^-\to\bar\nu_\mu\spac\mu^-$ requires dealing with a next-to-leading power problem in soft-collinear effective theory, in which endpoint-divergent convolution integrals arise. The operator basis for the decay amplitude contains operators in which the light spectator anti-quark in the $B$-meson is treated as a soft field and those in which it is treated as a collinear field. The endpoint divergences arise in the soft-collinear overlap region. They can be removed using the refactorization-based subtraction (RBS) scheme developed in \cite{Liu:2019oav,Liu:2020wbn,Beneke:2020ibj}. To this end, one introduces a scale parameter $\Lambda$ such that
\begin{equation}
   \Lambda_{\rm QCD} \ll \Lambda \ll m_B \,,
\end{equation}
which separates the genuinely soft region from collinear modes with large light-cone momentum \cite{Cornella:2022ubo,Cornella:2026lkp}. The soft overlap contributions in the endpoint-divergent collinear integrals are then removed and absorbed into corresponding soft matrix elements. In this process, the QED-generalized decay constants $F_\pm$ get modified and become $\Lambda$-dependent parameters, defined as
\begin{equation}\label{eq:Fdeffinal}
\begin{aligned}
   & \frac{1}{\myR}\,
    \langle\spac 0\spac|\spac\bar u\spac\bigg[ 1 
    - \theta\bigg(\! - \Lambda - \frac{in\cdot\!\overleftarrow{D}}{n\cdot v} \bigg) \bigg]\spac
    \Gamma\spac b_v\spac Y_\nb^{(M_2)\dagger}\spac|B^-(v)\rangle \\
   &= \frac{i}{2}\spac\sqrt{m_B}\,\,
    \mathrm{Tr}\left\{ \left[ \frac{\nbsl}{2\spac\nb\cdot v}\,F_-(\Lambda)
    + \frac{\nsl}{2\spac n\cdot v}\,F_+(\Lambda) \right] 
    \Gamma\spac\frac{1+\slashed{v}}{2}\spac\gamma_5\spac \right\} .
\end{aligned}
\end{equation}
The $\theta$-function ensures that momentum modes of the spectator anti-quark with large negative light-cone momentum $n\cdot p_u<-\Lambda\spac n\cdot v$ are excluded from the soft matrix element. With this subtraction, one finds that the operator on the left-hand side of the equation can now be renormalized multiplicatively. The renormalized parameters $F_\pm(\Lambda,\mu)$ obey the evolution equations \cite{Cornella:2022ubo,Cornella:2026lkp}
\begin{equation}\label{eq:RGEsforF}
   \frac{d}{d\ln\mu}\,F_\mp(\Lambda,\mu) 
   = - \gamma_{F_\mp}(\Lambda,\mu)\,F_\mp(\Lambda,\mu) \,,
\end{equation}
with the one-loop anomalous dimensions 
\begin{equation}\label{eq:RGEsforFsimp}
   \gamma_{F_\mp}(\Lambda,\mu) 
   = \gamma_{\rm hl}(\alpha_s)  
    + \frac{Q_u\spac\alpha}{4\pi} \left[ - (4\pm 2)\,Q_{M_2} - 3\spac Q_u 
    - 2\spac Q_{M_2}\spac\ln\frac{\mu^2}{\Lambda^2} \right] 
    + \mathcal{O}(\alpha\spac\alpha_s) \,,
\end{equation}
where $\gamma_{\rm hl}=-\alpha_s/\pi+\order{\alpha_s^2}$ is the well-known QCD anomalous dimension of heavy-light current operators in HQET \cite{Broadhurst:1991fz},
$Q_{M_2}=-1$ and $Q_u=2/3$. The dependence on $\Lambda$ in this result indicates that the QED-generalized decay constants contain Sudakov double logarithms of the cutoff parameter~$\Lambda$.

\subsection{QED-generalized distribution amplitudes}

Motivated by the above, one may consider gauge-invariant non-local currents of the form 
\begin{equation}
   \bar u(z\spac n)[zn,0]\spac\Gamma\spac b_v\spac Y_\nb^{(M_2)\dagger}(0) \,,
\end{equation} 
where the soft spectator field is delocalized along the $n^\mu$ direction, opposite to the direction $\nb^\mu$ of the charged final-state particle. The $B$-meson matrix element of this current defines the QED-generalized LCDAs $\Phi_\pm$ via \cite{Beneke:2022msp}\footnote{Our definition matches that of $\Phi_{B,\otimes}$ for $\otimes=(0,-)$ in this paper.} 
\begin{equation}\label{eq:LCDAbardef}
\begin{aligned}
   &\frac{1}{\myR}\,
    \langle\spac 0\spac|\spac\bar u\,\delta\bigg(\om-\frac{in\cdot\!\overleftarrow D}{n\cdot v}\bigg)\spac
    \Gamma \spac b_v\spac Y_{\nb}^{(M_2)\dagger}\spac|B^-(v)\rangle \\
   &= \frac{i}{2} \sqrt{m_B}\,F_{\rm QCD}\,\,
    \mathrm{Tr}\left\{ \left[ \frac{\nbsl}{2\spac\nb\cdot v}\,\Phi_+(\om)
    + \frac{\nsl}{2\spac n\cdot v}\,\Phi_-(\om) \right] 
    \Gamma\spac\frac{1+\slashed{v}}{2}\spac\gamma_5\spac \right\} ,
   \end{aligned}
\end{equation}
which generalizes the definition \eqref{eq:nLCDA} valid in pure QCD. It follows that
\begin{align}\label{eq:order_alpha_neg}
   \Phi_\pm(\om) = \phi_\pm(\om) + \order{\alpha} \,.
\end{align}
Once QED effects are included, the analytic structure of the LCDAs is modified, and their support extends over the entire real axis \cite{Beneke:2022msp}. QED evolution generates a non-vanishing contribution for negative $\om$ values starting at $\order{\alpha}$, even if the distribution is initially supported only for positive $\om$. Following the treatment in \cite{Beneke:2022msp}, by pulling out $F_{\rm QCD}$ one absorbs all QED effects in the LCDAs, leading to the normalization condition \begin{equation}\label{eq:LCDA_norm}
   \int_{-\infty}^\infty\!d\om\,\Phi_{\pm}(\om) 
   = 1 + \mathcal{O}(\alpha)
\end{equation}
for the dimensionally regularized (bare) LCDAs.

The definitions \eqref{eq:Fdeffinal} and \eqref{eq:LCDAbardef} allow us to establish an exact connection between the QED-generalized decay constants and LCDAs, which reads
\begin{equation}\label{eq:FtoPhi}
   \frac{d}{d\spac\Lambda}\,\frac{F_\mp(\Lambda)}{F_{\rm QCD}}
   = \Phi_\pm(\om) \Big|_{\om=-\Lambda} \,.
\end{equation}
This relation is the main result of our paper. The key point is that the dependence of $F_\mp(\Lambda)$ on the cutoff can be calculated as long as $\Lambda\gg\Lambda_{\rm QCD}$ is in the perturbative domain. This allows us to obtain a model-independent prediction for the asymptotic negative tail of the QED-generalized LCDAs. Moreover, we can use the RG evolution equation \eqref{eq:RGEsforF} to derive an equation controlling the scale dependence of the negative tails
\begin{align}\label{eq:RG_negtail}
   \frac{d}{d\ln\mu}\,\Phi_\pm(\om,\mu) \underset{\om\to -\infty}{=}
   Q_{M_2}\spac Q_u\spac\frac{\alpha}{\pi\spac\om} + \order{\alpha_s\spac\alpha,\alpha^2} .
\end{align}
We will come back to this relation in Section~\ref{sec:RG}.

\begin{figure}
\centering
\includegraphics[height=2.5cm]{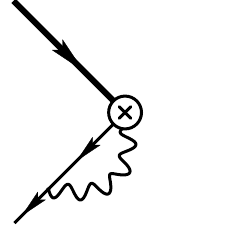} 
\includegraphics[height=2.5cm]{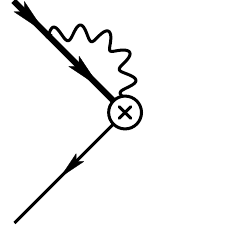} 
\includegraphics[height=2.5cm]{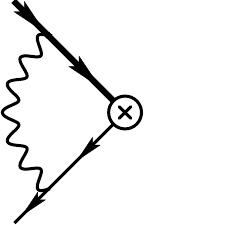} 
\hspace{-3mm}
\includegraphics[height=2.5cm]{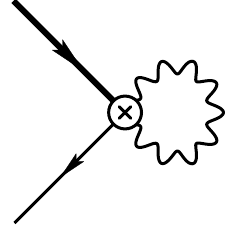} 
\hspace{3mm}
\includegraphics[height=2.5cm]{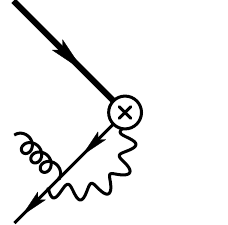} 
\includegraphics[height=2.5cm]{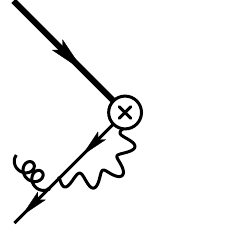}
\hspace{-1cm}
\caption{One-loop diagrams contributing to the perturbative calculation of the cutoff dependence of the matrix element \eqref{eq:Fdeffinal} in the region $\Lambda\gg\Lambda_\mathrm{QCD}$. Thick (thin) lines represent the heavy (light) quark. The last two graphs yield the term proportional to the three-particle LCDA $\phi_{3g}$ in \eqref{eq:Lambda_evolution}. Diagrams in which a gluon is emitted from the current (not shown) give vanishing contributions.} 
\label{fig:thetaops}
\end{figure}

Explicit expressions for the scale dependence of the QED-generalized decay constants have been given in (4.90) of \cite{Cornella:2026lkp}. They read 
\begin{equation}\label{eq:Lambda_evolution}
\begin{aligned}
   \frac{d}{d\ln\Lambda}\,\frac{F_-(\Lambda,\mu)}{F_{\rm QCD}(\mu)}
   &= - Q_{M_2}\spac Q_u\,\frac{\alpha}{2\pi}\,\bigg[ 
    \int_0^\infty\!d\omega\,\phi_-(\omega,\mu) 
    \left( \ln\frac{\mu^2}{\Lambda\spac\omega} + 1 \right) \\
   &\qquad - 2 \int_0^\infty\!d\omega \int_0^\infty\!d\omega_g\,\phi_{3g}(\omega,\omega_g,\mu) 
    \left( \frac{1}{\omega_g}\,\ln\frac{\omega+\omega_g}{\omega} 
    - \frac{1}{\omega+\omega_g} \right) \bigg] \,, \\
   \frac{d}{d\ln\Lambda}\,\,\frac{F_+(\Lambda,\mu)}{F_{\rm QCD}(\mu)}
   &= - Q_{M_2}\spac Q_u\,\frac{\alpha}{2\pi} \int_0^\infty\!d\omega\,
    \phi_+(\omega,\mu)\spac\ln\frac{\mu^2}{\Lambda\spac\omega} \,.
\end{aligned}
\end{equation} 
These results were derived by calculating the one-loop diagrams shown in Figure~\ref{fig:thetaops} assuming $\Lambda\gg\Lambda_\mathrm{QCD}$. The right-hand side involves integrals over the LCDAs $\phi_\pm$ and $\phi_{3g}$ defined in pure QCD, which is consistent since we work to first order in the QED coupling $\alpha$. The fact that only these three LCDAs appear follows from the structure of the operator basis derived in \cite{Cornella:2026lkp}. Using \eqref{eq:FtoPhi}, these relations directly translate into the asymptotic behavior of the LCDAs in the region of large negative momentum. We find
\begin{equation}\label{eq:QEDtail}
\begin{aligned}
   \Phi_+(\om,\mu) &\underset{\om\to -\infty}{=} 
    \frac{\alpha}{3\pi}\,\frac{1}{(-\om)} 
    \left[ \ln\frac{\mu^2}{(-\om)\spac\om_-}
    + 1 - 2\spac\lambda_{3g} + \order{\frac{\Lambda_{\rm QCD}}{\om}} \right] 
    + \order{\alpha_s\spac\alpha,\alpha^2} , \\
   \Phi_-(\om,\mu) &\underset{\om\to -\infty}{=} 
    \frac{\alpha}{3\pi}\,\frac{1}{(-\om)} 
    \left[ \ln\frac{\mu^2}{(-\om)\spac\om_+}
    + \order{\frac{\Lambda_{\rm QCD}}{\om}} \right] 
    + \order{\alpha_s\spac\alpha,\alpha^2} .
\end{aligned}
\end{equation}
This asymptotic expansion features the non-perturbative parameters $\om_{\pm}$ and $\lambda_{3g}$ defined by
\begin{equation}\label{eq:moments}
\begin{aligned}
   \ln\frac{\om_\pm(\mu)}{\nu} 
   &= \int_0^\infty\!d\om\,\phi_\pm(\om,\mu)\spac\ln\frac{\om}{\nu} \,, \\
   \lambda_{3g}(\mu) 
   &= \int_0^\infty\!d\omega \int_0^\infty\!d\omega_g\,\phi_{3g}(\omega,\omega_g,\mu)
    \left( \frac{1}{\omega_g}\spac\ln\frac{\omega+\omega_g}{\omega} - \frac{1}{\omega+\omega_g} \right) ,
\end{aligned}
\end{equation}
where $\nu$ is an arbitrary scale introduced to form a dimensionless ratio for the argument of the logarithms. At the accuracy of \eqref{eq:QEDtail}, the moments in \eqref{eq:moments} are evaluated at leading order in $\alpha_s$, so that the right-hand side of the first relation is well defined. The integral defining $\lambda_{3g}$ would be well defined even if the radiative $\order{\alpha_s}$ tails of the LCDAs are included. 

The expressions in \eqref{eq:QEDtail} provide model-independent predictions for the asymptotic tails of the QED-generalized LCDAs in the region $\om\ll-\Lambda_{\rm QCD}$. They independently confirm the existence of the negative support of the LCDAs $\Phi_\pm(\omega)$ first identified in \cite{Beneke:2022msp}. Our findings prove that these tails are present for any value of the renormalization scale $\mu$. Comparing the first relation in \eqref{eq:QEDtail} to the result for the asymptotic behavior at large $\omega$ values of the function $\phi_+(\omega)$ defined in pure QCD, we observe a similar leading $1/\om$ and $\ln(\omega/\mu)/\om$ behavior. However, in contrast to this result, non-perturbative quantities already appear at leading power in the QED-induced tail. Moreover, the  determination of these parameters involves non-local objects rather than matrix elements of local operators. The reason lies in the fact that the presence of the light-like Wilson line in \eqref{eq:LCDAbardef} is a source of non-locality which does not disappear for $|\omega|\gg\Lambda_{\rm QCD}$, even though the quark current becomes local in this limit.

\section{Renormalization-group evolution}
\label{sec:RG}

The RG evolution equation for the leading QED-generalized LCDA $\Phi_+(\omega,\mu)$ has been worked out in \cite{Beneke:2022msp}. Using the notations of this paper, the scale dependence at negative $\omega$ values takes the form (with $\Phi_+^<(\om,\mu)\equiv\Phi_+(\om,\mu)$ restricted to $\om<0$)
\begin{equation}\label{eq:RG_Phibp}
   \frac{d}{d\ln\mu}\,\Phi_+^<(\om,\mu)
   = - \int_{-\infty}^\infty\!d\om'\,\Gamma_{\Phi_+}^{<}(\om,\om',\mu)\,\Phi_+(\om',\mu) \,.
\end{equation}
Using the fact that $\Phi_\pm(\omega',\mu)=\order{\alpha}$ for $\omega'<0$, and dropping terms of $\order{\alpha^2}$, we obtain
\begin{equation}\label{eq:RG_alpha}
\begin{aligned}  
   \frac{d}{d\ln\mu}\,\Phi_+^<(\om,\mu)
   &\underset{\omega<0}{=} \,\frac{2\alpha}{3\pi} 
    \int_0^\infty\!d\om'\,\frac{\phi_+(\om',\mu)}{\om'-\om} 
    - \frac{C_F\spac\alpha_s}{\pi} \left( \ln{\frac{\mu}{-\om} - \frac12} \right) 
    \Phi_+^<(\om,\mu) \\
   &\quad + \frac{C_F\spac\alpha_s}{\pi}\,\int_{-\infty}^\om\!d\om'\,
    \frac{1}{\om'\spac(\om-\om')}\,\Big[ (\om+\om')\,\Phi_+^<(\om',\mu) - 2\om\,
    \Phi_+^<(\om,\mu) \Big] \,.
\end{aligned}
\end{equation}
The first term shows that a negative tail is necessarily generated by RG evolution from the LCDA in the positive-momentum region. In the limit $\om\to -\infty$, and neglecting $\order{\alpha_s}$ effects, we recover the evolution equation \eqref{eq:RG_negtail} for the asymptotic tail derived in Section~\ref{sec:LCDA}, which provides a non-trivial consistency check of our formalism. Neglecting the scale dependence of $\phi_+$, the solution of \eqref{eq:RG_alpha} reads
\begin{equation}\label{eq:RG_alphasol}
   \Phi^<_+(\om,\mu) 
   = \Phi^<_+(\om,\mu_0) + \frac{2\alpha}{3\pi}\,\ln\frac{\mu}{\mu_0} 
    \int_0^\infty d\om'\,\frac{\phi_+(\om',\mu_0)}{\om'-\om} + \order{\alpha_s\spac\alpha,\alpha^2} \,,
\end{equation}
where the initial condition at the scale $\mu_0$ is constrained, for large negative $\om$, by the asymptotic expression derived in \eqref{eq:QEDtail}.

We can improve the above solution by resumming the QCD logarithms generated at one-loop order via the $\order{\alpha_s}$ terms in \eqref{eq:RG_alpha}. Following the procedure outlined in \cite{Lee:2005gza,Liu:2020eqe,Beneke:2022msp}, the solution of \eqref{eq:RG_alpha} can be obtained analytically in this approximation. We introduce the evolution functions
\begin{equation}\label{eq:adef}
\begin{aligned}
   a(\mu,\mu_0)
   &= - \int_{\mu_0}^\mu\!\frac{d\mu'}{\mu'}\,\frac{C_F\spac\alpha_s(\mu')}{\pi} \,, \\ 
   V(\mu,\mu_0)
   &= - \int_{\mu_0}^\mu\!\frac{d\mu'}{\mu'} \left[ \frac{C_F\spac\alpha_s(\mu')}{\pi}\spac
    \ln\frac{\mu'}{\mu_0} - \frac{C_F\spac\alpha_s(\mu')}{2\pi} \right] -\frac{Q_u^2\spac\alpha}{2\pi}\spac\ln^2\frac{\mu}{\mu_0},
\end{aligned}
\end{equation}
where $V(\mu,\mu_0)$ resums leading logarithms in both QCD and QED, together with partial next-to-leading logarithmic QCD contributions. The RG-improved expression for $\Phi^{<}_+$ can then be written in the compact form
\begin{equation}\label{eq:res<}
\begin{aligned}
   \Phi^<_+(\om,\mu)
   &= e^{V(\mu,\mu_0)+2\gamma_E\spac a} \int_0^\infty\!\frac{d\om'}{\om'} \left( \frac{\mu_0}{\om'} \right)^a 
    \bigg[ G_a^{2,0}\bigg(\frac{-\om}{\om'}\bigg)\,\Phi^<_+(-\om',\mu_0) \\
   &\hspace{5.74cm} - Q_{M_2}\spac Q_u\,\frac{\alpha}{\pi}\spac\ln\frac{\mu}{\mu_0}\,
    G_a^{2,1}\bigg(\frac{-\om}{\om'}\bigg)\,\phi_+(\om',\mu_0) \bigg] \,,
\end{aligned}
\end{equation}
where $a\equiv a(\mu,\mu_0)$. The integration kernels are expressed in terms of Meijer-G functions $G_a^{m,n}(\tau)$, whose definitions can be found in \cite{Liu:2020eqe}.

\section{Phenomenological model}
\label{sec:model}

We now construct a phenomenological model for the negative-support component of $\Phi_+(\omega)$. For the LCDAs in pure QCD, we adopt the exponential model functions \cite{Grozin:1996pq,Braun:2017liq} 
\begin{equation}\label{eq:LCDAs_modeling}
\begin{aligned}
   \phi_+(\omega,\mu_0) 
   &= \frac{\omega}{\omega_0^2}\,e^{-\frac{\omega}{\omega_0}} \,, \\
   \phi_{3g}(\omega,\omega_g,\mu_0) 
   &= \frac{\lambda_E^2-\lambda_H^2}{6\spac\omega_0^5}\,\omega\spac\omega_g\,
    e^{-\frac{\omega+\omega_g}{\omega_0}} \,,   
\end{aligned}
\end{equation}
valid at a low scale $\mu_0=1$\,GeV. The QCD equations of motion relate the two-particle LCDA $\phi_-$ to $\phi_+$ and the three-particle LCDA $\phi_{3g}$ \cite{Kawamura:2001jm,Braun:2017liq}. One obtains 
\begin{equation}
   \phi_-(\omega,\mu_0)
   = \frac{1}{\omega_0}\,e^{- \frac{\omega}{\omega_0}}
    \left[ 1 - \frac{\lambda_E^2-\lambda_H^2}{18\spac\omega_0^2}
    \left( 2 - \frac{4\spac\omega}{\omega_0} + \frac{\omega^2}{\omega_0^2} \right) \right] .
\end{equation}
Using these model functions, the hadronic parameters defined in \eqref{eq:moments} evaluate to \cite{Cornella:2026lkp}
\begin{equation} 
   \ln\frac{\omega_-(\mu_0)}{\omega_0}
   = \frac{\lambda_E^2-\lambda_H^2}{18\spac\omega_0^2} - \gamma_E \,, \qquad
   \lambda_{3g}(\mu_0)
   = \frac{\lambda_E^2-\lambda_H^2}{36\spac\omega_0^2} \,.
\end{equation}
The hadronic input parameters entering our model functions are listed in Table~\ref{tab:inputs}. We use the \texttt{RunDec} package \cite{Chetyrkin:2000yt} to perform the RG evolution of $\alpha_s(\mu)$ and take $\alpha(\mu_0)=1/134$ for the electromagnetic coupling at $\mu_0=1\,\mathrm{GeV}$. The value $\om_0=\lambda_B$ is fixed by the first inverse moment of $\phi_+$. Note that these model functions correspond to ``tree-level'' expressions for the LCDAs and neglect the perturbative radiative tails at large $\omega$ and $\omega_g$ \cite{Lange:2003ff,Lee:2005gza}. 

We now turn to the description of the negative-support LCDA $\Phi^<_+$. Motivated by the $\order{\alpha}$ solution~\eqref{eq:RG_alphasol}, we write 
\begin{equation}\label{eq:model}
   \Phi_+^<(\om,\mu) = \Phi_+^<(\om,\mu_0)
    + \frac{2\alpha}{3\pi}\,\Phi_+^{\rm mix}(\om,\mu_0)\spac\ln\frac{\mu}{\mu_0} \,,
\end{equation}
where the second term describes the radiative mixing with the positive-support LCDA $\phi_+$, while the first term represents the genuine non-perturbative contribution at the scale $\mu_0$. In order for the model to reproduce the asymptotic behavior shown in \eqref{eq:QEDtail}, these functions must satisfy 
\begin{equation}
\begin{aligned}
   \Phi_+^{\rm mix}(\om,\mu_0) 
   &\underset{\om\to-\infty}{\to} - \frac{1}{\om} \,, \\
   \Phi_+^<(\om,\mu_0)
   &\underset{\om\to-\infty}{\to} \Phi_+^{\rm tail}(\om,\mu_0) \,,
\end{aligned}
\end{equation}
where within our model the radiative tail reads 
\begin{equation}\label{eq:tail}
   \Phi_+^{\rm tail}(\om,\mu_0)
   = \frac{\alpha}{3\pi\spac(-\om)}
    \left( \ln\frac{\mu_0^2}{-\om\spac\om_0} + 1 + \gamma_E - \frac{\lambda_E^2-\lambda_H^2}{9\spac\omega_0^2} \right) .
\end{equation}
We model the mixing term using its $\mathcal{O}(\alpha)$ expression derived from~\eqref{eq:RG_alphasol}, which yields
\begin{equation}
   \Phi_+^{\rm mix}(\om,\mu_0)
   = \frac{1}{\om_0} \left[ 1 - \frac{\om}{\om_0}\,e^{-\frac{\om}{\om_0}}\,
    \mathrm{Ei}\bigg(\frac{\om}{\om_0}\bigg) \right] ,
\end{equation}
where $\mathrm{Ei}(x)$ denotes the exponential integral function
\begin{equation}
    \mathrm{Ei}(-x) = - \int_x^\infty\frac{dt}{t}\,e^{-t} \,; \quad x>0 \,.
\end{equation}
Following \cite{Lee:2005gza}, we choose the initial condition $\Phi_+^<(\omega,\mu_0)$ such that the asymptotic behavior \eqref{eq:QEDtail} is smoothly matched to the non-perturbative region via a step-like function,
\begin{equation}
   \Phi_+^<(\om,\mu_0) 
   = \left( 1 - e^{-\frac{\om^2}{\om_t^2}} \right) \Phi_+^{\rm tail}(\om,\mu_0) \,,
\end{equation}
where $\om_t\sim\rm\order{\Lambda_{QCD}}$ is a cutoff scale controlling the transition between the non-perturbative and asymptotic regions. For our numerical estimates, we set $\omega_t=\omega_0$. 

\begin{table}
\centering
\renewcommand{\arraystretch}{1.15}
\setlength{\tabcolsep}{6pt}
\begin{tabular}{c|c|c}
\hline\rowcolor{\shadecolor{20}}
Parameter & Value & Reference \\
\hline
$\alpha_s(m_Z)$ & 0.118 & \cite{ParticleDataGroup:2024cfk} \\ 
$\lambda_E^2$ & $0.03\,\mathrm{GeV}^2$ & \cite{Nishikawa:2011qk} \\
$\lambda_H^2$ & $0.06\,\mathrm{GeV}^2$ & \cite{Nishikawa:2011qk} \\
$\lambda_B$ & $0.383\,\mathrm{GeV}$ &\cite{Khodjamirian:2020hob} \\
\hline
\end{tabular}
\caption{\label{tab:inputs}
Numerical values of the model parameters.}
\end{table}

\begin{figure}[t]
\centering
\includegraphics[scale=0.92]{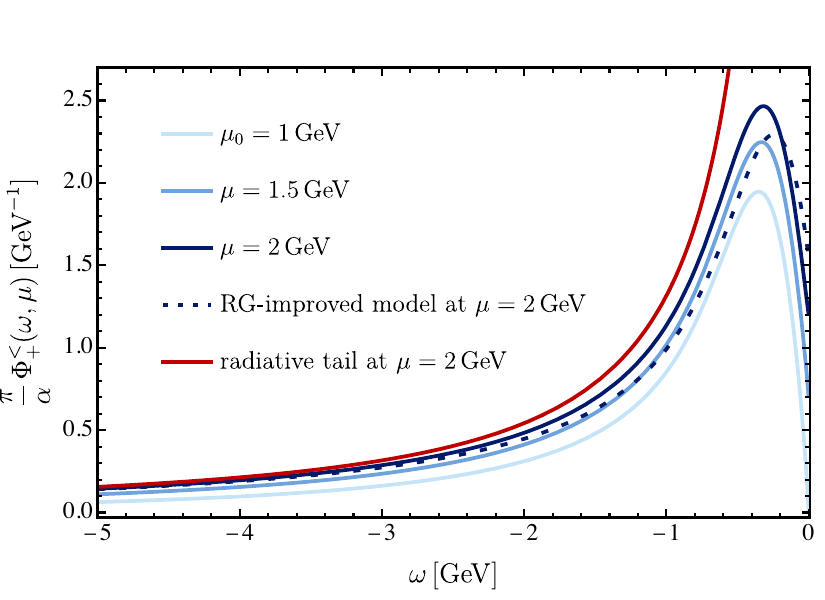}
\caption{Scale evolution of the model function \eqref{eq:model} for the negative-momentum component of the LCDA $\Phi_+(\om,\mu)$ (blue curves). The red curve shows the asymptotic form of the negative-momentum tail derived in \eqref{eq:tail}. The dashed blue curve shows the result at $\mu=2\,\mathrm{GeV}$ obtained by evolving the model from $\mu_0=1\,\mathrm{GeV}$ to $2\,\mathrm{GeV}$ via the resummed expression \eqref{eq:res<}.}
\label{fig:phiplus}
\end{figure}

We now have everything in place to present numerical estimates for the negative-momentum component of the QED-generalized LCDA $\Phi_+^<(\omega)$. Our results are collected in Figure~\ref{fig:phiplus}, where we plot the model function \eqref{eq:model} for $\Phi_+^<(\om,\mu)$ at different values of the renormalization scale $\mu$. Starting from a vanishing value at the origin at the initial scale $\mu_0$, $\Phi_+^<(\om,\mu)$ acquires a non-zero value there as $\mu$ increases \cite{Beneke:2020vnb}, signaling a violation of the pure-QCD scaling behavior~\eqref{eq:scaling} implied by conformal symmetry. For comparison, we also show the result at $\mu=2\, \mathrm{GeV}$ obtained by evolving the model~\eqref{eq:model} from $\mu_0=1\,\mathrm{GeV}$ using the RG evolution~\eqref{eq:res<}. The resulting curve closely follows the one obtained directly for the model function for values of $\om$ below $-1.5\,\mathrm{GeV}$, where the asymptotic behavior~\eqref{eq:tail} provides a reliable description, indicating approximate RG invariance of our parameterization in this region. However, the two curves gradually deviate from each other in the low negative-momentum region, which is not constrained by the asymptotic analysis. 

\section{Conclusion}
\label{sec:conclusion}

We have studied electromagnetic corrections to $B$-meson LCDAs in HQET, focusing on the emergence of a radiative tail at large negative light-cone momentum induced by soft-photon radiation. We have shown that this negative-momentum region is directly connected to QED-generalized decay constants involving soft Wilson lines associated with charged final-state particles. Using this framework, we have derived a model-independent expression for the asymptotic behavior of the QED-generalized LCDAs in the region $\om\ll-\Lambda_{\rm QCD}$. In contrast to the QCD radiative tail at large positive momentum, the QED-induced tail already depends at leading power on non-perturbative quantities, namely integrals over the  two- and three-particle $B$-meson LCDAs defined in pure QCD. We have further investigated the RG evolution of the generalized LCDAs and obtained resummed expressions valid at leading order in RG-improved perturbation theory. Finally, we have proposed a phenomenological model for the negative-momentum component of the leading-power LCDA $\Phi_+(\omega,\mu)$, which is consistent with the asymptotic constraints and RG evolution derived in this work. Our results provide a systematic framework for incorporating soft QED effects into heavy-meson LCDAs and may be relevant for future precision studies of exclusive $B$-meson decays.

\section*{Acknowledgments}

This research has received funding from the Cluster of Excellence PRISMA${}^{++}$ (EXC 2118/2, Project ID 390831469) funded by the German Research Foundation (DFG) within the Germany Excellence Strategy, and from the European Research Council (ERC) under the European Union’s Horizon 2022 Research and Innovation Program (ERC Advanced Grant agreement No.~101097780, EFT4jets). Views and opinions expressed in this work are those of the authors only and do not necessarily reflect those of the European Union or the European Research Council Executive Agency. Neither the European Union nor the granting authority can be held responsible for them. 

\bibliographystyle{JHEP}
\bibliography{refs}

\end{document}